\documentclass[aip,jcp,reprint,amsmath,amssymb,floatfix,citeautoscript,nofootinbib]{revtex4-2}
\usepackage{cancel}
\usepackage{amsmath}
\usepackage{physics}
\usepackage{graphicx}
\usepackage{amssymb}
\usepackage{amsthm}
\usepackage{bm}

\usepackage{booktabs}
\usepackage{dcolumn}
\newcolumntype{x}[1]{D{.}{.}{#1}}

\usepackage{ragged2e}
\usepackage{txfonts}
\allowdisplaybreaks

\usepackage{lipsum}

\usepackage[usenames,dvipsnames]{xcolor}
\usepackage[breaklinks=true,colorlinks=true,linkcolor=blue,urlcolor=blue,citecolor=blue]{hyperref}

\newcommand{\si}{Supplementary Material}  

\usepackage{tabularx}

\begin{document}

\title{
Adsorption energies and decomposition barrier heights for ethylene carbonate on the surface of lithium from cluster-based quantum chemistry
}

\author{Ethan A. Vo}
\author{Hung T. Vuong}
\author{Zachary K. Goldsmith}
\affiliation{Department of Chemistry, Columbia University, New York, NY 10027 USA}
\author{Hong-Zhou Ye}
\affiliation{Department of Chemistry and Biochemistry, University of Maryland, College Park, MD, 20742, USA}
\affiliation{Institute for Physical Science and Technology, University of Maryland, College Park, MD, 20742, USA}
\author{Yujing Wei}
\author{Sohang Kundu}
\author{Ardavan Farahvash}
\affiliation{Department of Chemistry, Columbia University, New York, NY 10027 USA}
\author{Garvit Agarwal}
\affiliation{Schr{\"o}dinger, Inc., New York, NY 10036, USA}
\author{Richard A. Friesner}
\email{raf8@columbia.edu}
\affiliation{Department of Chemistry, Columbia University, New York, NY 10027 USA}
\author{Timothy C. Berkelbach}
\email{t.berkelbach@columbia.edu}
\affiliation{Department of Chemistry, Columbia University, New York, NY 10027 USA}
\affiliation{Initiative for Computational Catalysis, Flatiron Institute, New York, NY 10010 USA}

\begin{abstract}
For ethylene carbonate on the (100) surface of lithium, we calculate the adsorption energy in two binding motifs as well as the barrier height for a ring-opening decomposition reaction. 
We validate a scheme for producing results in the thermodynamic limit by correcting results obtained on finite lithium clusters containing only 40--100 atoms, which enables the use of hybrid density functionals, the random-phase approximation, and correlated wavefunction theories such as coupled-cluster theory and auxiliary-field quantum Monte Carlo.
We find that the high-level theories agree to within 2--5~kcal/mol and can therefore serve as benchmarks for more affordable methods.
Using our reference data, we demonstrate that generalized gradient approximation functionals, such as PBE, are not sufficiently accurate for reaction barrier heights, and we identify $\omega$B97X-V as an especially promising functional for the interfacial chemistry of electrolyte solvents at lithium metal anodes.
\end{abstract}

\maketitle

\section{Introduction}

Density functional theory (DFT) has been widely used to study surface
chemistry, such as the initial stages of solid-electrolyte interphase formation
at the surface of lithium metal
anodes~\cite{Wang2001,Wang2002,Wang2005,Ebadi2016}. However, the accuracy of
DFT depends on the employed exchange-correlation functional, and achieving high
accuracy on several properties simultaneously---such as surface energies,
adsorption energies, and barrier heights---is
challenging~\cite{Schimka2010,Schmidt2018}.  For example, many functionals have
been evaluated on several benchmark datasets of experimental adsorption
energies (CE39~\cite{Wellendorff2015} and ADS41~\cite{MallikarjunSharada2019}) and reaction barrier
heights (SBH10~\cite{MallikarjunSharada2017} and SBH17~\cite{Tchakoua2022}),
finding that no single functional achieves errors below 5~kcal/mol (about
0.2~eV).  Perhaps surprisingly, the PBE functional~\cite{Perdew1996}, a generalized gradient
approximation (GGA), was one of the best performing on those datasets, yielding
mean absolute errors of 7.8~kcal/mol (0.34~eV) for adsorption
energies~\cite{MallikarjunSharada2019} and 2.8~kcal/mol (0.12~eV) on barrier
heights~\cite{Tchakoua2022}. More complicated functionals, such as meta-GGAs or
hybrids, do not reliably show improved performance on these databases.
However, these databases primarily consist of small molecules (four atoms or
less) on transition-metal surfaces, and the transferability of these
conclusions to other types of surface chemistry is unclear.

In a previous work from our groups, we studied the reaction energetics of
ethylene carbonate (EC) decomposition in the presence of a single lithium
atom~\cite{Debnath2023}, which is a model for reductive decomposition in the presence of a lithium ion in solution. By performing accurate quantum chemistry calculations
using coupled-cluster (CC) theory and phaseless auxiliary-field quantum Monte Carlo
(AFQMC), we were able to benchmark a large number of commonly used functionals.
As expected for gas-phase chemistry, GGAs such as PBE exhibited large errors,
underpredicting decomposition barrier heights by about 10~kcal/mol (0.43~eV).
Here, we aim to provide similar benchmark data for EC on the surface of lithium
metal.  
These interactions are potentially of great interest for understanding the formation of the solid-electrolyte interphase at lithium anodes in lithium-ion batteries.  
Benchmark quantum chemical data is useful, for example, to train a machine-learning potential for molecular dynamics simulations of a realistic lithium-ion battery model, along the lines of recent efforts from our groups~\cite{Kundu2025,Stevenson2025,Wei2026}. 
Despite advances in periodic, wavefunction-based quantum
chemistry~\cite{Pisani2008,Gruber2018}, such calculations are still expensive,
especially for metals~\cite{Neufeld2022,Masios2023,Neufeld2023}.  Therefore, we
take a cluster-based approach, motivated in part by previous efforts to correct
periodic GGA calculations with more complicated functionals or higher levels of
theory~\cite{Hu2007,Ren2009,Boese2013,Bernard2015,Araujo2022}.

The simple idea is to perform calculations on model surfaces made of a finite
number of metal atoms comprising a cluster. Although passivation and
electrostatic embedding methods have been successfully applied to ionic
solids~\cite{Boese2013,Bernard2015,Shi2025}, their optimal construction for
metals is unclear, and so all of our clusters will be unpassivated.  For
sufficiently large clusters, we expect that calculated predictions
of energy differences $\Delta E = E_\text{B}-E_\text{A}$ will
converge to the correct, thermodynamic limit, but the rate of convergence may
be slow.

Whereas low levels of theory such as GGAs can be applied to clusters large
enough to observe convergence, high levels of theory typically cannot (as we
show below).  In this case, 
we can hope that the convergence behavior is similar between low and high
levels of theory and apply a simple finite-size correction,
\begin{subequations}
\label{eq:composite}
\begin{align}
\Delta E^\text{HL}(\infty) &\approx 
\Delta E^\text{HL}(N) + \Delta\Delta E^{\text{LL}}_\text{FS}(N) \\
\Delta\Delta E^\text{LL}_\text{FS}(N) &= \Delta E^\text{LL}(\infty) - \Delta E^\text{LL}(N),
\end{align}
\end{subequations}
where HL and LL are high-level and low-level theories, $N$
is the number of atoms in the metal cluster,
and $\Delta\Delta E^\text{LL}_\text{FS}(N)$ is a finite-size correction
for clusters containing $N$ atoms, computed with the low-level theory.
This strategy
only works if the two methods have a small non-parallelity error, 
i.e., $\Delta E^\text{HL}(N) - \Delta E^\text{LL}(N) \approx \text{constant}$,
for sufficiently large $N$.
In principle, the converged low-level energy differences $\Delta E^\text{LL}(\infty)$ can be
obtained from a large but finite cluster or from a large supercell with
periodic boundary conditions (PBCs).  In the latter case,
Eqs.~(\ref{eq:composite}) can be simply reinterpreted as a correction to the
low-level periodic energy,
\begin{subequations}
\begin{align}
\Delta E^\text{HL}(\text{PBC}) &\approx 
\Delta E^\text{LL}(\text{PBC}) + \Delta\Delta E^{\text{HL-LL}}(N) \\
\Delta\Delta E^\text{HL-LL}(N) &= \Delta E^\text{HL}(N) - \Delta E^\text{LL}(N),
\end{align}
\end{subequations}
which is how it has been applied in the 
literature for solids~\cite{Hu2007,Ren2009,Sauer2019,Araujo2022,Sheldon2024}
and, recently, liquids~\cite{O’Neill2025}.

\section{Methods}

\begin{figure}[t]
    \centering
    \includegraphics[scale=1.0]{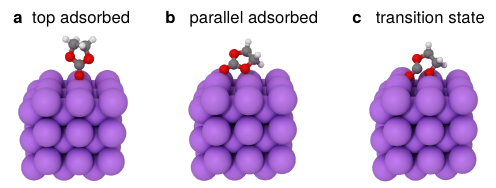}
    \caption{
    Geometries of ethylene carbonate on lithium for the top adsorbed structure (a),
    the parallel adsorbed structure (b), and the transition state to ring opening
    from the parallel structure (c).
    }
    \label{fig:geoms}
\end{figure}

We study the behavior of EC on the (100) surface of lithium with a lattice
constant of 3.44~$\AA$.  We aim to accurately calculate the adsorption energy
$\Delta E_\text{top/par}$ in the so-called top and parallel
geometries~\cite{Ebadi2016} and the reaction barrier height $\Delta E^\ddagger$
for EC ring-opening~\cite{Agarwal2026} from the more stable parallel geometry.
The geometries of the three needed structures are shown in
Fig.~\ref{fig:geoms}, where optimization was done with PBCs 
and a $3\times 3\times 6$ slab containing 54 lithium atoms, as
reported in Ref.~\onlinecite{Agarwal2026}.
To construct larger models from which to extract clusters, we embedded the geometry
of each periodic EC+slab model into the geometry of a clean, geometry-optimized lithium surface.

We tested two methods for extracting clusters. In the first, we extract hemispherical clusters
by keeping all lithium atoms within a given distance of the EC (almost all clusters generated
this way are slightly asymmetric). In the second, we extract parallelepiped slabs using the lattice vectors of BCC lithium.
As shown in the Supplementary Material, we observe qualitatively similar results with both cluster shapes, but the hemispheres give access to more clusters, and so we use them throughout this work.
We only consider clusters with an even number of lithium atoms,
ensuring a closed-shell electronic configuration, and all calculations are
performed for the ground-state singlet using spin-restricted methods (exploratory
calculations found that spin-unrestricted calculations did not break spin symmetry).
Adsorption energies are calculated using counterpoise corrections to reduce
basis set superposition error.

As AFQMC is significantly more expensive to run than the other methods used in this work, we follow a systematic, cost-saving protocol that we have previously used to compare AFQMC and other high-level methods over large data sets (with applications to small organolithium complexes~\cite{Debnath2023}, organic systems for which the available experimental thermochemistry data disagreed with the best high level theoretical calculations~\cite{Wei2024}, and atomization energies~\cite{Shee2018}, bond dissociation energies~\cite{Shee2019}, ligand dissociation energies~\cite{Rudshteyn2020}, ionization potentials of transition metal containing molecules~\cite{Rudshteyn2022, Vuong2026}).
Because the cost of AFQMC depends on the number of determinants used in the trial function, we first determine a standardized, relatively inexpensive trial function that yields agreement to within 2--3 kcal/mol for most of the cases in the data set (for organic molecule thermochemistry, such a single determinant trial is often sufficient). 
We then identify all of the outliers and attempt to refine them by improving the trial function. 
For cases containing a single transition metal that do not have too great of a multiconfigurational character, we have developed an optimized heat-bath CI (HCI) trial, which generally provides very good results~\cite{Vuong2026}, but for more challenging cases, a much longer determinantal expansion based on a CCSD reference wavefunction~\cite{Mahajan2025} provides consistently superior results~\cite{Vuong2026}. 

Hartree-Fock (HF), DFT, random-phase approximation (RPA), and CCSD calculations were performed with PySCF~\cite{Sun2017,Sun2020}.
CCSD calculations were performed in a truncated virtual basis using frozen natural orbitals calculated with second-order perturbation theory and a conservative truncation threshold.
Some DFT calculations and all DLPNO-CCSD(T)~\cite{Riplinger2013,Riplinger2013a} calculations were performed with ORCA~6.0~\cite{Neese2011,Neese2025}, using NormalPNO settings and the (T0) approximation.
AFQMC calculations were performed with an in-house code~\cite{Shee2018} using the phaseless approximation~\cite{Zhang2003} with localized orbitals~\cite{Weber2022}.
As discussed above, most of our AFQMC results use a spin-restricted single-determinant trial wavefunction with DFT orbitals obtained using the B3LYP functional, although we use an HCI trial wavefunction for a few outliers (see below).
Full details of our AFQMC calculations are given in the \si.
Core orbitals were frozen in all correlated calculations.

\section{Results}

\begin{figure}
    \centering
    \includegraphics[scale=1.0]{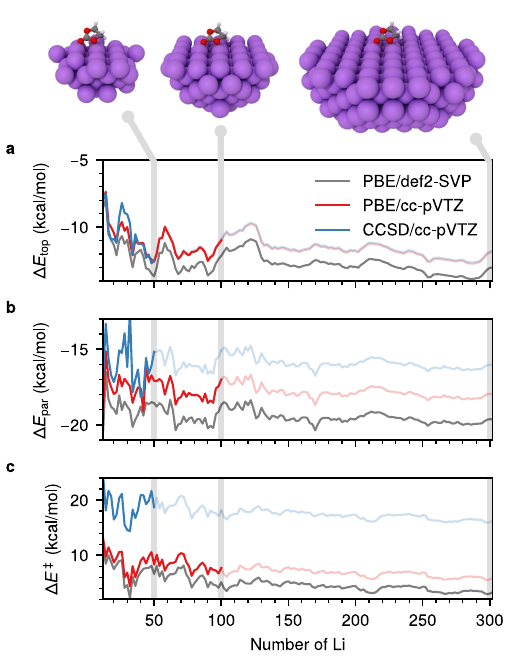}
    \caption{
    Adsorption energy in the top (a) and parallel (b) geometries, and decomposition reaction barrier height (c)
    for EC on hemispherical clusters of increasing size. 
    Results are shown for the methods and basis sets indicated in the legend, and faint lines
    indicate approximate extensions of the data based on the PBE/def2-SVP results.
    Geometries at top show the parallel adsorption geometry with clusters containing 
    50, 100, and 300 lithium atoms.
    }
    \label{fig:pbe_ccsd}
\end{figure}

First, to gain insight into the expected convergence of the adsorption energies
and barrier heights, we perform calculations using PBE in the def2-SVP basis
set, a relatively small double-zeta basis. The low cost of these calculations
allows us to go to larger clusters than is possible with larger basis sets
and higher levels of theory. In Fig.~\ref{fig:pbe_ccsd}, we show results for
clusters containing up to 300 lithium atoms.  The size dependence of the energy
differences is rather erratic, which we attribute to the asymmetric addition of lithium atoms as the cluster is grown.  For the adsorption energies $\Delta
E_\text{top/par}$, the convergence is relatively quick: beyond 50 atoms in the
cluster, almost all predictions are within about 1~kcal/mol of the value at the
largest cluster size.  The convergence of the barrier height $\Delta
E^\ddagger$ is slower, and the predicted value is not accurate to 1--2~kcal/mol
until there are more than 100 lithium atoms in the cluster. 

Next, we repeat PBE calculations using the larger cc-pVTZ basis set, which is more
appropriate for correlated levels of theory, for clusters containing up to 100
lithium atoms; results are shown in Fig.~\ref{fig:pbe_ccsd}. Importantly, we
see that the fluctuations with cluster size are very correlated across basis sets.  
To quantify these correlations, we use the non-parallelity error (NPE), which
is commonly used for scans of a molecular potential energy surface. Here, we
define the mean-absolute-deviation NPE between two levels of theory (e.g., HL and LL),
\begin{equation}
\text{NPE} = \frac{1}{N_\text{c}} \sum_{N_\text{Li}} 
    \left| \Delta\Delta E^\text{HL-LL}(N_\text{Li}) - \overline{\Delta\Delta E^\text{HL-LL}} \right|,
\end{equation}
where $N_\text{Li}$ is the number of lithium atoms in the cluster and
$N_\text{c}$ is the number of clusters over which we average (this definition
of the NPE is less sensitive to outliers than a max-min definition).  In
Tab.~\ref{tab:npe}, we present the NPE between PBE/cc-pVTZ and PBE/def2-SVP
values, which is less than 0.4~kcal/mol for all three energy differences. If we
assume this small NPE holds for larger clusters, we can safely extend the PBE/cc-pVTZ
data, as shown graphically in Fig.~\ref{fig:pbe_ccsd}. With this extension, we
predict converged values given in Tab.~\ref{tab:results}.
On the basis of these PBE results, we can calculate finite-size corrections
$\Delta \Delta E^\text{PBE}_\text{FS}(N_\text{Li})$ 
which will be applied to the correlated levels of theory.
For example, $\Delta \Delta E^\text{PBE}_\text{FS}(N_\text{Li}=100) = -0.9, -0.9, -2.1$~kcal/mol
for the top adsorption energy, parallel adsorption energy, and barrier height, respectively.
Importantly, our PBE results (which use large but finite clusters and a Gaussian basis set) agree with those of Ref.~\onlinecite{Agarwal2026} (which use periodic boundary conditions and a plane-wave basis set) to within about 1~kcal/mol, and thus either one could be reasonably used to define the finite-size corrections (the small difference may be attributed to periodic interactions and pseudopotential effects).

\begin{table}[t]
    \begin{ruledtabular}
        \begin{tabular}{lddd}
            Method & \multicolumn{1}{c}{$\text{NPE}_\text{top}$} & \multicolumn{1}{c}{$\text{NPE}_\text{par}$} & \multicolumn{1}{c}{$\text{NPE}^\ddagger$} \\
            \midrule
            PBE           &  0.13 &  0.18 & 0.34 \\
            RPA@PBE       &  2.78 &  2.39 & 6.73 \\
            RPA@HF        &  0.74 &  1.34 & 2.25 \\
            CCSD          &  0.60 &  0.84 & 1.17 \\
            DLPNO-CCSD(T) &  0.85 &  1.08 & 2.54 \\
            AFQMC         &  1.03 &  1.62 & 1.80 \\
            \hline
            PBE-D3        &  0.08 &  0.35 & 0.31 \\
            PBE0          &  0.29 &  0.20 & 0.47 \\
            PBE0-D3       &  0.25 &  0.43 & 0.47 \\
            B3LYP         &  0.43 &  0.33 & 0.60 \\
            B3LYP-D3      &  0.34 &  0.36 & 0.64 \\
            $\omega$B97X-V&  0.92 &  0.88 & 1.71 \\
            $\omega$B97X-D3BJ&  1.04 &  0.88 & 2.26 \\
        \end{tabular}
    \end{ruledtabular}
    \caption{Non-parallelity errors (NPE) for adsorption energies and barrier
    heights (kcal/mol) in the cc-pVTZ basis with respect to PBE/def2-SVP. Averaging was performed over clusters containing
more than 10 lithiums and less than 100 lithiums for PBE and RPA, less than 50 lithiums for CCSD, and less than 40 lithiums for DLPNO-CCSD(T) and AFQMC.} 
    \label{tab:npe}
\end{table}

\begin{table}[b]
    \begin{ruledtabular}
        \begin{tabular}{lddd}
            Method & \multicolumn{1}{c}{$\Delta E_\text{ads}^\text{top}$} & \multicolumn{1}{c}{$\Delta E_\text{ads}^\text{par}$} & \multicolumn{1}{c}{$\Delta E^\ddagger$} \\
            \midrule
            PBE (PBC/PW)~\cite{Agarwal2026}  & -10.9 & -17.2 &  6.9 \\
            PBE (def2-SVP)& -13.0 & -19.6 &  3.0 \\
            PBE           & -11.9 & -17.9 &  5.6 \\
            RPA@PBE       & -11.4 & -14.2 &  9.1 \\
            RPA@HF        & -13.2 & -17.6 & 18.3 \\
            CCSD          & -11.8 & -16.0 & 16.0 \\
            DLPNO-CCSD(T) & -12.4 & -19.1 & 15.5 \\
            AFQMC         &  -9.1 & -21.4 & 17.4 \\
            Best estimate &  -11.1 \pm 1.4 & -18.8 \pm 2.2 & 16.3 \pm 0.8 \\
            \hline
            PBE-D3        & -14.1 & -22.7 & 4.6 \\
            PBE0          & -11.3 & -17.2 & 13.7 \\
            PBE0-D3       & -13.6 & -22.1 & 12.5 \\
            B3LYP         & -10.5 & -14.6 &  6.9 \\
            B3LYP-D3      & -13.4 & -21.6 &  4.7 \\
            $\omega$B97X-V& -12.1 & -20.8 & 17.9  \\
            $\omega$B97X-D3BJ& -11.9 & -22.2 & 20.8 \\
        \end{tabular}
    \end{ruledtabular}
    \caption{Estimated adsorption energies and barrier heights (kcal/mol) in the dilute, thermodynamic limit for the methods used in this work. AFQMC results are obtained with a spin-restricted, single-determinant trial wavefunction with B3LYP orbitals, except for four adsorption energy data points, which were obtained with a more accurate HCI trial (see text). The ``best estimate'' predictions are obtained from the mean and standard deviation of the CCSD, DLPNO-CCSD(T), and AFQMC predictions. Except where indicated, calculations are performed with the cc-pVTZ basis set.}
    \label{tab:results}
\end{table}

\begin{table}[b]
    \begin{ruledtabular}
        \begin{tabular}{ldddd}
            & \multicolumn{4}{c}{$\Delta E_\mathrm{ads}^\mathrm{par}(N_\mathrm{Li})$} \\
            Method & \multicolumn{1}{c}{$N_\mathrm{Li}=26$} & \multicolumn{1}{c}{$N_\mathrm{Li}=28$} & \multicolumn{1}{c}{$N_\mathrm{Li}=38$} & \multicolumn{1}{c}{$N_\mathrm{Li}=40$} \\
            \midrule
            CCSD          & -15.3 & -13.9 & -16.3 & -18.2 \\
            DLPNO-CCSD(T) & -17.6 & -16.0 & -20.1 & -18.8 \\
            AFQMC@B3LYP   & -23.3 & -14.5 & -22.4 & -26.2 \\
            AFQMC@HCI     & -18.3 & -13.8 & -17.8 & -21.0 
        \end{tabular}
    \end{ruledtabular}
    \caption{Parallel adsorption energies (kcal/mol) for the four cluster sizes indicated, showing improved agreement between CC theory and AFQMC with a multideterminental HCI trial wavefunction. All AFQMC results have a statistical uncertainty of about 1~kcal/mol.}
    \label{tab:hci}
\end{table}

We performed analogous calculations using CCSD in the cc-pVTZ basis set, and results are also shown in Fig.~\ref{fig:pbe_ccsd} for clusters containing up to 50 lithium atoms (1736 basis functions).
Again the fluctuations are quite strongly correlated with those of PBE, and the NPEs given in Tab.~\ref{tab:npe} are around 1~kcal/mol. 
Using PBE data to extend our CCSD results, as shown in Fig.~\ref{fig:pbe_ccsd}---or equivalently, applying the finite-size corrections $\Delta\Delta E^\text{PBE}_\text{FS}(N_\text{Li}=50)$---yields the large-cluster predictions given in Tab.~\ref{tab:results}.

\begin{figure*}[t]
    \centering
    \includegraphics[scale=1.0]{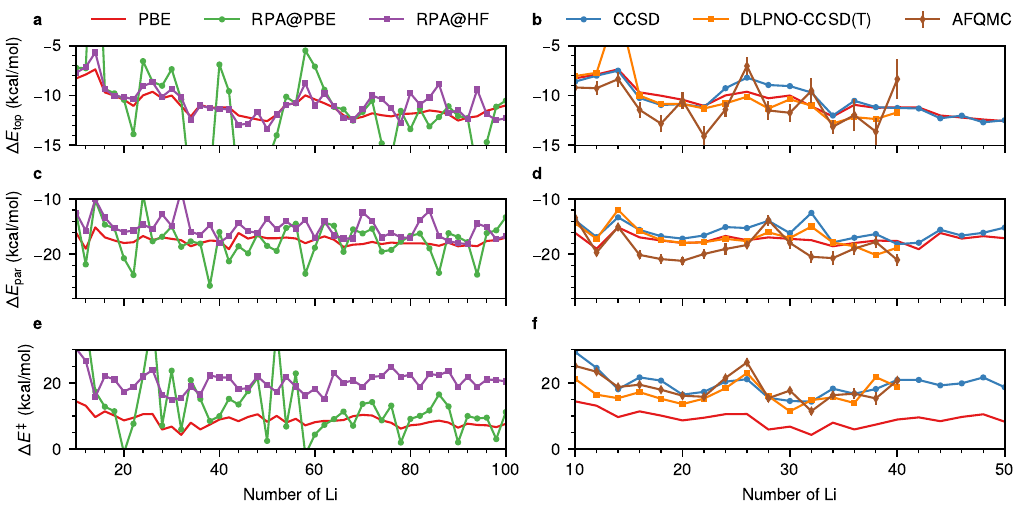}
    \caption{
    The same as in Fig.~\ref{fig:pbe_ccsd}, but for correlated methods indicated in the legend.  AFQMC results are shown with statistical error bars.
    }
    \label{fig:corr}
\end{figure*}

In Fig.~\ref{fig:corr}, we show results obtained using other correlated theories, including the RPA, DLPNO-CCSD(T), and AFQMC, where RPA is applied to clusters containing up to 100 atoms, and
DLPNO-CCSD(T) and AFQMC are applied to clusters containing up to 40 atoms, due
to their higher cost.  Of these methods, only RPA has seen much application to
surface science~\cite{Schimka2010,Schmidt2018,Oudot2024}, and it has been used
in cluster-based corrections of the type being explored
here~\cite{Hu2007,Ren2009,Sheldon2024}.  We present RPA results performed using PBE
orbitals and using HF orbitals; the former is significantly more common in the
community. 
Both RPA results are again quite erratic with
fluctuations of 10~kcal/mol or more. Our testing has suggested 
that these fluctuations are due to the nonlocal exact exchange term in the RPA
total energy. Of the two, we observe that RPA@HF has smaller fluctuations and
that they are more correlated with those of PBE (NPEs of about 1--2~kcal/mol),
which suggests that cluster-based corrections with PBE would be reliable.

We find that CCSD and DLPNO-CCSD(T) agree to better than 3~kcal/mol for almost all clusters studied (up to 40 lithium atoms).
However, we find that energy differences calculated with the DLPNO approximation deviate from their canonical counterparts by a few kcal/mol, and up to 5~kcal/mol for the barrier height (see \si). 
We attribute this behavior to the increasingly metallic character of the clusters and associated long-range electron correlation. 
Moreover, the applicability of CCSD(T) for the calculation of energy differences on metal surfaces is questionable, given its divergent correlation energy of bulk three-dimensional
metals~\cite{Masios2023,Neufeld2023} (in contrast, RPA and CCSD are well-behaved infinite-order perturbation theories for three-dimensional metals).

Turning to the comparison with AFQMC, we observe that for small clusters, a single-determinant trial appears to be sufficient to achieve 2--3 kcal/mol agreement with CCSD and CCSD(T).
However, for a few of the larger clusters, we found deviations as large as 7~kcal/mol.
As per our usual protocol, we repeated the AFQMC calculations with an improved HCI trial wavefunction~\cite{Wei2024}.
Specifically, for the clusters containing 26, 28, 38, and 40, we calculated the parallel adsorption energy using the HCI trial, and the comparison to results obtained with the single-determinant trial are given in Tab.~\ref{tab:hci}.
The agreement with CCSD and DLPNO-CCSD(T) is substantially improved, to within our targeted 2--3 kcal/mol accuracy threshold.
We conclude from this investigation that the results of CCSD, DLPNO-CCSD(T), and AFQMC provide reasonable reference energies for the systems studied in this paper. 

Finally, we apply the PBE-based finite-size corrections (based on results for clusters with 100 atoms for RPA, 50 atoms for CCSD, and 40 atoms for DLPNO-CCSD(T) and AFQMC) to produce our best predictions for all levels of theory given in Tab.~\ref{tab:results} (however, we note that most of the correlated methods have NPEs of about 1--2.5~kcal/mol, as reported in Tab.~\ref{tab:npe}).
We find that CCSD, DLPNO-CCSD(T), and AFQMC agree to within about 5~kcal/mol for most energy differences.
Lacking any convincing evidence to prefer one over the others, we obtain our overall best estimates by averaging these three levels of theory.
Given the uncertainties associated with PBE-based finite-size corrections, we consider these best estimates to provide accurate benchmark values.
Importantly, the associated uncertainties are small enough to evaluate the performance of common density functionals.

\begin{figure}[b]
    \centering
    \includegraphics[scale=1.0]{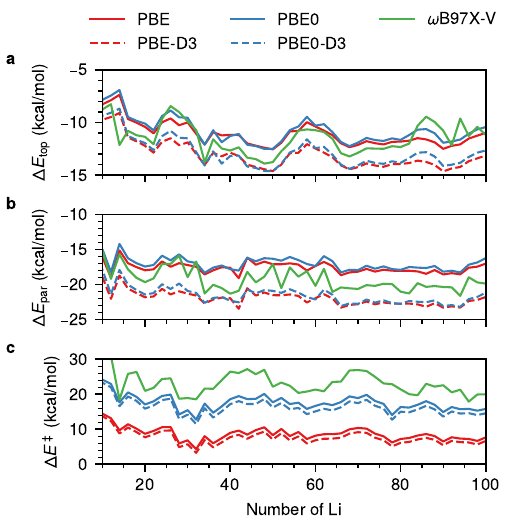}
    \caption{
    The same is in Figs.~\ref{fig:pbe_ccsd} and \ref{fig:corr}, but for the various
    density functionals indicated.
    }
    \label{fig:dft}
\end{figure}

In Fig.~\ref{fig:dft}, we show the performance of several functionals, with and without dispersion corrections: PBE, PBE-D3~\cite{Grimme2010}, PBE0 (a hybrid)~\cite{Adamo1999}, PBE0-D3, and the range-separated hybrids $\omega$B97X-V~\cite{Mardirossian2014} and $\omega$B97X-D3BJ~\cite{Najibi2018} (for most functionals tested, the differences between results with D3 and VV10~\cite{Vydrov2010} dispersion corrections are usually less than 1~kcal/mol).
The NPE is clearly small for all functionals tested, and we report the NPE values in Tab.~\ref{tab:npe}. Given the low NPE, we confidently apply the PBE finite-size corrections (based on the other DFT results on clusters with 100 atoms) to obtain the final predictions in Tab.~\ref{tab:results}, which we discuss next in our conclusion. 

\section{Conclusion}

Based on the comparison in Tab.~\ref{tab:results}, we conclude that dispersion
corrections make a significant impact on adsorption energies (increasing the
magnitude by 2--5 kcal/mol) but a negligible impact on barrier heights.
Conversely, hybrid functionals exhibit adsorption energies that are similar to their GGA
counterparts, but much larger barrier heights; compared to the PBE barrier
height of 5.6~kcal/mol, the PBE0 and $\omega$B97X-V barrier heights are 13.7
and 17.9~kcal/mol. These larger barrier heights are in good agreement with our theoretical best estimate.

We propose that $\omega$B97X-V is an accurate functional for lithium-electrolyte surface chemistry, agreeing with our theoretical best estimates to within 2~kcal/mol, which is within the error bar of our approximations.
With a lower cost for the dispersion correction, $\omega$B97X-D3BJ performs only slightly worse.
The hybrid PBE0 (and PBE0-D3) performs quite well and better than B3LYP, especially for the barrier height.
In contrast, GGAs like PBE and PBE-D3 predict reasonable adsorption energies, but underpredict the reaction barrier height by more than 10~kcal/mol, in agreement with our previous findings on EC decomposition in the presence of a single lithium atom~\cite{Debnath2023}. 
These results conclusively demonstrate that the good performance of PBE on the SBH10 and SBH17 datasets is not transferable to all kinds of metal surface chemistry.

Looking forward, the low NPE of hybrid functionals with respect to PBE suggests that cluster-based corrections or other types of finite-size corrections can be safely used for accurate metal surface chemistry as long as uncertainties of about 2--3~kcal/mol are tolerable. 
This finding is practically important given the relatively limited availability and high cost of hybrid functionals for periodic calculations on metals---in contrast to GGA-based calculations, which are now routine. 
Through either direct hybrid calculations or cluster-based corrections, we anticipate the production of high-quality reference data against which modern machine-learning interatomic potentials can be trained for molecular dynamics studies of EC decomposition and subsequent SEI formation.

\section*{Supplementary material}

See the supplementary material for comparison of parallelepiped and hemisphere clusters, comparison of canonical CC with DLPNO CC, and additional details on AFQMC calculations.

\section*{Acknowledgements}

This work was supported by the Columbia Center for Computational Electrochemistry.
We acknowledge computing resources from Columbia University's
Shared Research Computing Facility project, which is supported by NIH Research
Facility Improvement Grant 1G20RR030893-01, and associated funds from the New
York State Empire State Development, Division of Science Technology and
Innovation (NYSTAR) Contract C090171, both awarded April 15, 2010.
This research used resources of the Oak Ridge Leadership Computing Facility at
the Oak Ridge National Laboratory, which is supported by the Office of Science
of the U.S. Department of Energy under Contract DE-AC05-00OR22725.
The Flatiron Institute is a division of the Simons Foundation.

\section*{Data availability statement}
The data that support the findings of this study are available from the
corresponding author upon reasonable request.

\bibliography{cluster,afqmc_refs}

@Article{Mahajan2025,
  author    = {Mahajan, Ankit and Thorpe, James H. and Kurian, Jo S. and Reichman, David R. and Matthews, Devin A. and Sharma, Sandeep},
  journal   = {J. Chem. Theory Comput.},
  title     = {Beyond CCSD(T) Accuracy at Lower Scaling with Auxiliary Field Quantum Monte Carlo},
  year      = {2025},
  issn      = {1549-9618},
  month     = feb,
  number    = {4},
  pages     = {1626--1642},
  volume    = {21},
  comment   = {doi: 10.1021/acs.jctc.4c01314},
  doi       = {10.1021/acs.jctc.4c01314},
  publisher = {American Chemical Society},
  url       = {https://doi.org/10.1021/acs.jctc.4c01314},
}

@Article{Vuong2026,
  author    = {Vuong, Hung T. and Mahajan, Ankit and Weber, John L. and Shee, James and Reichman, David R. and Friesner, Richard A.},
  journal   = {J. Chem. Theory Comput.},
  title     = {Evaluating Multiconfigurational Trials for Accurate Phaseless Auxiliary-Field Quantum Monte Carlo on 3d Transition Metal Complexes},
  year      = {2026},
  issn      = {1549-9618},
  month     = mar,
  pages     = {Article ASAP},
  comment   = {doi: 10.1021/acs.jctc.5c01936},
  doi       = {10.1021/acs.jctc.5c01936},
  publisher = {American Chemical Society},
  url       = {https://doi.org/10.1021/acs.jctc.5c01936},
}

@Article{Wei2024,
  author    = {Wei, Yujing and Debnath, Sibali and Weber, John L. and Mahajan, Ankit and Reichman, David R. and Friesner, Richard A.},
  journal   = {J. Phys. Chem. A},
  title     = {Scalable Ab Initio Electronic Structure Methods with Near Chemical Accuracy for Main Group Chemistry},
  year      = {2024},
  issn      = {1089-5639},
  month     = jul,
  number    = {28},
  pages     = {5796--5807},
  volume    = {128},
  comment   = {doi: 10.1021/acs.jpca.4c02853},
  doi       = {10.1021/acs.jpca.4c02853},
  publisher = {American Chemical Society},
  url       = {https://doi.org/10.1021/acs.jpca.4c02853},
}

@Article{Zhang2003,
  author    = {Zhang, Shiwei and Krakauer, Henry},
  journal   = {Phys. Rev. Lett.},
  title     = {Quantum Monte Carlo Method using Phase-Free Random Walks with Slater Determinants},
  year      = {2003},
  month     = apr,
  number    = {13},
  pages     = {136401},
  volume    = {90},
  doi       = {10.1103/PhysRevLett.90.136401},
  publisher = {American Physical Society},
  refid     = {10.1103/PhysRevLett.90.136401},
  url       = {https://link.aps.org/doi/10.1103/PhysRevLett.90.136401},
}

@Article{Shee2018,
  author    = {Shee, James and Arthur, Evan J. and Zhang, Shiwei and Reichman, David R. and Friesner, Richard A.},
  journal   = {J. Chem. Theory Comput.},
  title     = {Phaseless Auxiliary-Field Quantum Monte Carlo on Graphical Processing Units},
  year      = {2018},
  issn      = {1549-9618},
  month     = aug,
  number    = {8},
  pages     = {4109--4121},
  volume    = {14},
  comment   = {doi: 10.1021/acs.jctc.8b00342},
  doi       = {10.1021/acs.jctc.8b00342},
  publisher = {American Chemical Society},
  url       = {https://doi.org/10.1021/acs.jctc.8b00342},
}

@Article{Shee2019,
  author    = {Shee, James and Rudshteyn, Benjamin and Arthur, Evan J. and Zhang, Shiwei and Reichman, David R. and Friesner, Richard A.},
  journal   = {J. Chem. Theory Comput.},
  title     = {On Achieving High Accuracy in Quantum Chemical Calculations of 3d Transition Metal-Containing Systems: A Comparison of Auxiliary-Field Quantum Monte Carlo with Coupled Cluster, Density Functional Theory, and Experiment for Diatomic Molecules},
  year      = {2019},
  issn      = {1549-9618},
  month     = apr,
  number    = {4},
  pages     = {2346--2358},
  volume    = {15},
  comment   = {doi: 10.1021/acs.jctc.9b00083},
  doi       = {10.1021/acs.jctc.9b00083},
  publisher = {American Chemical Society},
  url       = {https://doi.org/10.1021/acs.jctc.9b00083},
}

@Article{Rudshteyn2020,
  author    = {Rudshteyn, Benjamin and Coskun, Dilek and Weber, John L. and Arthur, Evan J. and Zhang, Shiwei and Reichman, David R. and Friesner, Richard A. and Shee, James},
  journal   = {J. Chem. Theory Comput.},
  title     = {Predicting Ligand-Dissociation Energies of 3d Coordination Complexes with Auxiliary-Field Quantum Monte Carlo},
  year      = {2020},
  issn      = {1549-9618},
  month     = may,
  number    = {5},
  pages     = {3041--3054},
  volume    = {16},
  comment   = {doi: 10.1021/acs.jctc.0c00070},
  doi       = {10.1021/acs.jctc.0c00070},
  publisher = {American Chemical Society},
  url       = {https://doi.org/10.1021/acs.jctc.0c00070},
}

@Article{Rudshteyn2022,
  author    = {Rudshteyn, Benjamin and Weber, John L. and Coskun, Dilek and Devlaminck, Pierre A. and Zhang, Shiwei and Reichman, David R. and Shee, James and Friesner, Richard A.},
  journal   = {J. Chem. Theory Comput.},
  title     = {Calculation of Metallocene Ionization Potentials via Auxiliary Field Quantum Monte Carlo: Toward Benchmark Quantum Chemistry for Transition Metals},
  year      = {2022},
  issn      = {1549-9618},
  month     = may,
  number    = {5},
  pages     = {2845--2862},
  volume    = {18},
  comment   = {doi: 10.1021/acs.jctc.1c01071},
  doi       = {10.1021/acs.jctc.1c01071},
  publisher = {American Chemical Society},
  url       = {https://doi.org/10.1021/acs.jctc.1c01071},
}

@Article{Weber2022,
  author    = {Weber, John L. and Vuong, Hung and Devlaminck, Pierre A. and Shee, James and Lee, Joonho and Reichman, David R. and Friesner, Richard A.},
  journal   = {J. Chem. Theory Comput.},
  title     = {A Localized-Orbital Energy Evaluation for Auxiliary-Field Quantum Monte Carlo},
  year      = {2022},
  issn      = {1549-9618},
  month     = jun,
  number    = {6},
  pages     = {3447--3459},
  volume    = {18},
  comment   = {doi: 10.1021/acs.jctc.2c00111},
  doi       = {10.1021/acs.jctc.2c00111},
  publisher = {American Chemical Society},
  url       = {https://doi.org/10.1021/acs.jctc.2c00111},
}

@Article{Ren2009,
  author    = {Ren, Xinguo and Rinke, Patrick and Scheffler, Matthias},
  journal   = {Phys. Rev. B},
  title     = {Exploring the random phase approximation: Application to {CO} adsorbed on {Cu(111)}},
  year      = {2009},
  month     = jul,
  pages     = {045402},
  volume    = {80},
  doi       = {10.1103/PhysRevB.80.045402},
  issue     = {4},
  numpages  = {8},
  publisher = {American Physical Society},
}

@Article{Hu2007,
  author    = {Hu, Qing-Miao and Reuter, Karsten and Scheffler, Matthias},
  journal   = {Phys. Rev. Lett.},
  title     = {Towards an Exact Treatment of Exchange and Correlation in Materials: Application to the ``{CO} Adsorption Puzzle'' and Other Systems},
  year      = {2007},
  month     = apr,
  pages     = {176103},
  volume    = {98},
  doi       = {10.1103/PhysRevLett.98.176103},
  issue     = {17},
  numpages  = {4},
  publisher = {American Physical Society},
}

@Article{Ebadi2016,
  author    = {Ebadi, Mahsa and Brandell, Daniel and Araujo, C. Moyses},
  journal   = {J. Chem. Phys.},
  title     = {Electrolyte decomposition on {Li}-metal surfaces from first-principles theory},
  year      = {2016},
  issn      = {1089-7690},
  month     = nov,
  number    = {20},
  pages     = {204701},
  volume    = {145},
  doi       = {10.1063/1.4967810},
  publisher = {AIP Publishing},
}

@Article{Perdew1996,
  author    = {Perdew, John P. and Burke, Kieron and Ernzerhof, Matthias},
  journal   = {Phys. Rev. Lett.},
  title     = {Generalized Gradient Approximation Made Simple},
  year      = {1996},
  issn      = {1079-7114},
  month     = oct,
  number    = {18},
  pages     = {3865--3868},
  volume    = {77},
  doi       = {10.1103/physrevlett.77.3865},
  publisher = {American Physical Society (APS)},
}

@Article{Oudot2024,
  author    = {Oudot, B. and Doblhoff-Dier, K.},
  journal   = {J. Chem. Phys.},
  title     = {Reaction barriers at metal surfaces computed using the random phase approximation: Can we beat {DFT} in the generalized gradient approximation?},
  year      = {2024},
  issn      = {1089-7690},
  month     = aug,
  number    = {5},
  pages     = {054708},
  volume    = {161},
  doi       = {10.1063/5.0220465},
  publisher = {AIP Publishing},
}

@Article{Wellendorff2015,
  author    = {Wellendorff, Jess and Silbaugh, Trent L. and Garcia-Pintos, Delfina and Nørskov, Jens K. and Bligaard, Thomas and Studt, Felix and Campbell, Charles T.},
  journal   = {Surf. Sci.},
  title     = {A benchmark database for adsorption bond energies to transition metal surfaces and comparison to selected DFT functionals},
  year      = {2015},
  issn      = {0039-6028},
  month     = oct,
  pages     = {36--44},
  volume    = {640},
  doi       = {10.1016/j.susc.2015.03.023},
  publisher = {Elsevier BV},
}

@Article{Neufeld2022,
  author    = {Neufeld, Verena A. and Ye, Hong-Zhou and Berkelbach, Timothy C.},
  journal   = {J. Phys. Chem. Lett.},
  title     = {Ground-State Properties of Metallic Solids from Ab Initio Coupled-Cluster Theory},
  year      = {2022},
  issn      = {1948-7185},
  month     = aug,
  number    = {32},
  pages     = {7497--7503},
  volume    = {13},
  doi       = {10.1021/acs.jpclett.2c01828},
  publisher = {American Chemical Society (ACS)},
}

@Article{Grimme2010,
  author    = {Grimme, Stefan and Antony, Jens and Ehrlich, Stephan and Krieg, Helge},
  journal   = {J. Chem. Phys.},
  title     = {A consistent and accurate ab initio parametrization of density functional dispersion correction ({DFT-D}) for the 94 elements {H-Pu}},
  year      = {2010},
  issn      = {1089-7690},
  month     = apr,
  number    = {15},
  pages     = {154104},
  volume    = {132},
  doi       = {10.1063/1.3382344},
  publisher = {AIP Publishing},
}

@Article{Wang2001,
  author    = {Wang, Yixuan and Nakamura, Shinichiro and Ue, Makoto and Balbuena, Perla B.},
  journal   = {J. Am. Chem. Soc.},
  title     = {Theoretical Studies To Understand Surface Chemistry on Carbon Anodes for Lithium-Ion Batteries: Reduction Mechanisms of Ethylene Carbonate},
  year      = {2001},
  issn      = {1520-5126},
  month     = nov,
  number    = {47},
  pages     = {11708--11718},
  volume    = {123},
  doi       = {10.1021/ja0164529},
  publisher = {American Chemical Society (ACS)},
}

@Article{Wang2002,
  author    = {Wang, Yixuan and Balbuena, Perla B.},
  journal   = {J. Phys. Chem. B},
  title     = {Theoretical Insights into the Reductive Decompositions of Propylene Carbonate and Vinylene Carbonate: Density Functional Theory Studies},
  year      = {2002},
  issn      = {1520-5207},
  month     = apr,
  number    = {17},
  pages     = {4486--4495},
  volume    = {106},
  doi       = {10.1021/jp014371t},
  publisher = {American Chemical Society (ACS)},
}

@Article{Wang2005,
  author    = {Wang, Yixuan and Balbuena, Perla B.},
  journal   = {Int. J. Quantum Chem.},
  title     = {Theoretical studies on cosolvation of {Li} ion and solvent reductive decomposition in binary mixtures of aliphatic carbonates},
  year      = {2005},
  issn      = {1097-461X},
  month     = jan,
  number    = {5},
  pages     = {724--733},
  volume    = {102},
  doi       = {10.1002/qua.20466},
  publisher = {Wiley},
}

@Article{Araujo2022,
  author    = {Araujo, Rafael B. and Rodrigues, Gabriel L. S. and dos Santos, Egon Campos and Pettersson, Lars G. M.},
  journal   = {Nat. Commun.},
  title     = {Adsorption energies on transition metal surfaces: towards an accurate and balanced description},
  year      = {2022},
  issn      = {2041-1723},
  month     = nov,
  number    = {1},
  pages     = {6853},
  volume    = {13},
  doi       = {10.1038/s41467-022-34507-y},
  publisher = {Springer Science and Business Media LLC},
}

@Article{Sun2020,
  author    = {Sun, Qiming and Zhang, Xing and Banerjee, Samragni and Bao, Peng and Barbry, Marc and Blunt, Nick S. and Bogdanov, Nikolay A. and Booth, George H. and Chen, Jia and Cui, Zhi-Hao and Eriksen, Janus J. and Gao, Yang and Guo, Sheng and Hermann, Jan and Hermes, Matthew R. and Koh, Kevin and Koval, Peter and Lehtola, Susi and Li, Zhendong and Liu, Junzi and Mardirossian, Narbe and McClain, James D. and Motta, Mario and Mussard, Bastien and Pham, Hung Q. and Pulkin, Artem and Purwanto, Wirawan and Robinson, Paul J. and Ronca, Enrico and Sayfutyarova, Elvira R. and Scheurer, Maximilian and Schurkus, Henry F. and Smith, James E. T. and Sun, Chong and Sun, Shi-Ning and Upadhyay, Shiv and Wagner, Lucas K. and Wang, Xiao and White, Alec and Whitfield, James Daniel and Williamson, Mark J. and Wouters, Sebastian and Yang, Jun and Yu, Jason M. and Zhu, Tianyu and Berkelbach, Timothy C. and Sharma, Sandeep and Sokolov, Alexander Yu. and Chan, Garnet Kin-Lic},
  journal   = {J. Chem. Phys.},
  title     = {Recent developments in the {PySCF} program package},
  year      = {2020},
  issn      = {1089-7690},
  month     = jul,
  number    = {2},
  pages     = {024109},
  volume    = {153},
  doi       = {10.1063/5.0006074},
  publisher = {AIP Publishing},
}

@Article{Schimka2010,
  author    = {Schimka, L. and Harl, J. and Stroppa, A. and Grüneis, A. and Marsman, M. and Mittendorfer, F. and Kresse, G.},
  journal   = {Nat. Mater.},
  title     = {Accurate surface and adsorption energies from many-body perturbation theory},
  year      = {2010},
  issn      = {1476-4660},
  month     = jul,
  number    = {9},
  pages     = {741--744},
  volume    = {9},
  doi       = {10.1038/nmat2806},
  publisher = {Springer Science and Business Media LLC},
}

@Article{Schmidt2018,
  author    = {Schmidt, Per S. and Thygesen, Kristian S.},
  journal   = {J. Phys. Chem. C},
  title     = {Benchmark Database of Transition Metal Surface and Adsorption Energies from Many-Body Perturbation Theory},
  year      = {2018},
  issn      = {1932-7455},
  month     = feb,
  number    = {8},
  pages     = {4381--4390},
  volume    = {122},
  doi       = {10.1021/acs.jpcc.7b12258},
  publisher = {American Chemical Society (ACS)},
}

@Article{MallikarjunSharada2019,
  author    = {Mallikarjun Sharada, Shaama and Karlsson, Rasmus K. B. and Maimaiti, Yasheng and Voss, Johannes and Bligaard, Thomas},
  journal   = {Phys. Rev. B},
  title     = {Adsorption on transition metal surfaces: Transferability and accuracy of {DFT} using the {ADS41} dataset},
  year      = {2019},
  issn      = {2469-9969},
  month     = jul,
  number    = {3},
  pages     = {035439},
  volume    = {100},
  doi       = {10.1103/physrevb.100.035439},
  publisher = {American Physical Society (APS)},
}

@Article{MallikarjunSharada2017,
  author    = {Mallikarjun Sharada, Shaama and Bligaard, Thomas and Luntz, Alan C. and Kroes, Geert-Jan and Nørskov, Jens K.},
  journal   = {J. Phys. Chem. C},
  title     = {{SBH10}: A Benchmark Database of Barrier Heights on Transition Metal Surfaces},
  year      = {2017},
  issn      = {1932-7455},
  month     = sep,
  number    = {36},
  pages     = {19807--19815},
  volume    = {121},
  doi       = {10.1021/acs.jpcc.7b05677},
  publisher = {American Chemical Society (ACS)},
}

@Article{Tchakoua2022,
  author    = {Tchakoua, T. and Gerrits, N. and Smeets, E. W. F. and Kroes, G.-J.},
  journal   = {J. Chem. Theory Comput.},
  title     = {{SBH17}: Benchmark Database of Barrier Heights for Dissociative Chemisorption on Transition Metal Surfaces},
  year      = {2022},
  issn      = {1549-9626},
  month     = dec,
  number    = {1},
  pages     = {245--270},
  volume    = {19},
  doi       = {10.1021/acs.jctc.2c00824},
  publisher = {American Chemical Society (ACS)},
}

@Article{Debnath2023,
  author    = {Debnath, Sibali and Neufeld, Verena A. and Jacobson, Leif D. and Rudshteyn, Benjamin and Weber, John L. and Berkelbach, Timothy C. and Friesner, Richard A.},
  journal   = {J. Phys. Chem. A},
  title     = {Accurate Quantum Chemical Reaction Energies for Lithium-Mediated Electrolyte Decomposition and Evaluation of Density Functional Approximations},
  year      = {2023},
  issn      = {1520-5215},
  month     = oct,
  number    = {44},
  pages     = {9178--9184},
  volume    = {127},
  doi       = {10.1021/acs.jpca.3c04369},
  publisher = {American Chemical Society (ACS)},
}

@Article{Adamo1999,
  author    = {Adamo, Carlo and Barone, Vincenzo},
  journal   = {J. Chem. Phys.},
  title     = {Toward reliable density functional methods without adjustable parameters: The {PBE0} model},
  year      = {1999},
  issn      = {1089-7690},
  month     = apr,
  number    = {13},
  pages     = {6158--6170},
  volume    = {110},
  doi       = {10.1063/1.478522},
  publisher = {AIP Publishing},
}

@Article{Pisani2008,
  author    = {Pisani, Cesare and Maschio, Lorenzo and Casassa, Silvia and Halo, Migen and Schütz, Martin and Usvyat, Denis},
  journal   = {J. Comput. Chem.},
  title     = {Periodic local MP2 method for the study of electronic correlation in crystals: Theory and preliminary applications},
  year      = {2008},
  issn      = {1096-987X},
  month     = may,
  number    = {13},
  pages     = {2113--2124},
  volume    = {29},
  doi       = {10.1002/jcc.20975},
  publisher = {Wiley},
}

@Article{Gruber2018,
  author    = {Gruber, Thomas and Liao, Ke and Tsatsoulis, Theodoros and Hummel, Felix and Grüneis, Andreas},
  journal   = {Phys. Rev. X},
  title     = {Applying the Coupled-Cluster Ansatz to Solids and Surfaces in the Thermodynamic Limit},
  year      = {2018},
  issn      = {2160-3308},
  month     = may,
  number    = {2},
  pages     = {021043},
  volume    = {8},
  doi       = {10.1103/physrevx.8.021043},
  publisher = {American Physical Society (APS)},
}

@Article{Masios2023,
  author    = {Masios, Nikolaos and Irmler, Andreas and Schäfer, Tobias and Grüneis, Andreas},
  journal   = {Phys. Rev. Lett.},
  title     = {Averting the Infrared Catastrophe in the Gold Standard of Quantum Chemistry},
  year      = {2023},
  issn      = {1079-7114},
  month     = oct,
  number    = {18},
  pages     = {186401},
  volume    = {131},
  doi       = {10.1103/physrevlett.131.186401},
  publisher = {American Physical Society (APS)},
}

@Article{Neufeld2023,
  author    = {Neufeld, Verena A. and Berkelbach, Timothy C.},
  journal   = {Phys. Rev. Lett.},
  title     = {Highly Accurate Electronic Structure of Metallic Solids from Coupled-Cluster Theory with Nonperturbative Triple Excitations},
  year      = {2023},
  issn      = {1079-7114},
  month     = oct,
  number    = {18},
  pages     = {186402},
  volume    = {131},
  doi       = {10.1103/physrevlett.131.186402},
  publisher = {American Physical Society (APS)},
}

@Article{Bernard2015,
  author    = {Bernard, Eddy and Houriez, Céline and Mitrushchenkov, Alexander O. and Guitou, Marie and Chambaud, Gilberte},
  journal   = {J. Chem. Phys.},
  title     = {Physisorbed {H$_2$@Cu(100)} surface: Potential and spectroscopy},
  year      = {2015},
  issn      = {1089-7690},
  month     = feb,
  number    = {5},
  pages     = {054703},
  volume    = {142},
  doi       = {10.1063/1.4907013},
  publisher = {AIP Publishing},
}

@Article{Shi2025,
  author    = {Shi, Benjamin X. and Rosen, Andrew S. and Schäfer, Tobias and Grüneis, Andreas and Kapil, Venkat and Zen, Andrea and Michaelides, Angelos},
  journal   = {Nat. Chem.},
  title     = {An accurate and efficient framework for modelling the surface chemistry of ionic materials},
  year      = {2025},
  issn      = {1755-4349},
  month     = aug,
  number    = {11},
  pages     = {1688--1695},
  volume    = {17},
  doi       = {10.1038/s41557-025-01884-y},
  publisher = {Springer Science and Business Media LLC},
}

@Article{Sauer2019,
  author    = {Sauer, Joachim},
  journal   = {Acc. Chem. Res.},
  title     = {Ab Initio Calculations for Molecule–Surface Interactions with Chemical Accuracy},
  year      = {2019},
  issn      = {1520-4898},
  month     = nov,
  number    = {12},
  pages     = {3502--3510},
  volume    = {52},
  doi       = {10.1021/acs.accounts.9b00506},
  publisher = {American Chemical Society (ACS)},
}

@Article{Boese2013,
  author    = {Boese, A. Daniel and Sauer, Joachim},
  journal   = {Phys. Chem. Chem. Phys.},
  title     = {Accurate adsorption energies of small molecules on oxide surfaces: {CO–MgO(001)}},
  year      = {2013},
  issn      = {1463-9084},
  number    = {39},
  pages     = {16481},
  volume    = {15},
  doi       = {10.1039/c3cp52321g},
  publisher = {Royal Society of Chemistry (RSC)},
}

@Article{Mardirossian2014,
  author    = {Mardirossian, Narbe and Head-Gordon, Martin},
  journal   = {Phys. Chem. Chem. Phys.},
  title     = {{$\omega$B97X-V}: A 10-parameter, range-separated hybrid, generalized gradient approximation density functional with nonlocal correlation, designed by a survival-of-the-fittest strategy},
  year      = {2014},
  issn      = {1463-9084},
  number    = {21},
  pages     = {9904},
  volume    = {16},
  doi       = {10.1039/c3cp54374a},
  publisher = {Royal Society of Chemistry (RSC)},
}

@Article{Vydrov2010,
  author    = {Vydrov, Oleg A. and Van Voorhis, Troy},
  journal   = {J. Chem. Phys.},
  title     = {Nonlocal van der {Waals} density functional: The simpler the better},
  year      = {2010},
  issn      = {1089-7690},
  month     = dec,
  number    = {24},
  pages     = {244103},
  volume    = {133},
  doi       = {10.1063/1.3521275},
  publisher = {AIP Publishing},
}

@Article{O’Neill2025,
  author    = {O’Neill, Niamh and Shi, Benjamin X. and Baldwin, William J. and Witt, William C. and Csányi, Gábor and Gale, Julian D. and Michaelides, Angelos and Schran, Christoph},
  journal   = {J. Chem. Theory Comput.},
  title     = {Towards Routine Condensed Phase Simulations with Delta-Learned Coupled Cluster Accuracy: Application to Liquid Water},
  year      = {2025},
  issn      = {1549-9626},
  month     = nov,
  number    = {22},
  pages     = {11710--11720},
  volume    = {21},
  doi       = {10.1021/acs.jctc.5c01377},
  publisher = {American Chemical Society (ACS)},
}

@Article{Najibi2018,
  author    = {Najibi, Asim and Goerigk, Lars},
  journal   = {J. Chem. Theory Comput.},
  title     = {The Nonlocal Kernel in van der {Waals} Density Functionals as an Additive Correction: An Extensive Analysis with Special Emphasis on the {B97M-V} and {$\omega$B97M-V} Approaches},
  year      = {2018},
  issn      = {1549-9626},
  month     = oct,
  number    = {11},
  pages     = {5725--5738},
  volume    = {14},
  doi       = {10.1021/acs.jctc.8b00842},
  publisher = {American Chemical Society (ACS)},
}

@Article{Agarwal2026,
  author    = {Agarwal, Garvit and Brock, Casey N. and Guha, Rishabh D. and Rao, Karun K. and Stevenson, James M. and Tiwari, Subodh C. and Fonari, Alexandr and Jacobson, Leif D. and Kwak, H. Shaun and Halls, Mathew D.},
  journal   = {Comput. Mater. Sci.},
  title     = {Insights into electrolyte reactivity at the {Li} metal surface from density functional theory},
  year      = {2026},
  issn      = {0927-0256},
  month     = jan,
  pages     = {114278},
  volume    = {261},
  doi       = {10.1016/j.commatsci.2025.114278},
  publisher = {Elsevier BV},
}

@Article{Sun2017,
  author    = {Sun, Qiming and Berkelbach, Timothy C. and Blunt, Nick S. and Booth, George H. and Guo, Sheng and Li, Zhendong and Liu, Junzi and McClain, James D. and Sayfutyarova, Elvira R. and Sharma, Sandeep and Wouters, Sebastian and Chan, Garnet Kin‐Lic},
  journal   = {WIREs Comput. Mol. Sci.},
  title     = {{PySCF}: the {Python}‐based simulations of chemistry framework},
  year      = {2017},
  issn      = {1759-0884},
  month     = sep,
  number    = {1},
  pages     = {e1340},
  volume    = {8},
  doi       = {10.1002/wcms.1340},
  publisher = {Wiley},
}

@Article{Neese2011,
  author    = {Neese, Frank},
  journal   = {WIREs Comput. Mol. Sci.},
  title     = {The {ORCA} program system},
  year      = {2011},
  issn      = {1759-0884},
  month     = jun,
  number    = {1},
  pages     = {73--78},
  volume    = {2},
  doi       = {10.1002/wcms.81},
  publisher = {Wiley},
}

@Article{Neese2025,
  author    = {Neese, Frank},
  journal   = {WIREs Comput. Mol. Sci.},
  title     = {Software Update: The {ORCA} Program System—Version 6.0},
  year      = {2025},
  issn      = {1759-0884},
  month     = mar,
  number    = {2},
  pages     = {e70019},
  volume    = {15},
  doi       = {10.1002/wcms.70019},
  publisher = {Wiley},
}

@Article{Riplinger2013,
  author    = {Riplinger, Christoph and Neese, Frank},
  journal   = {J. Chem. Phys.},
  title     = {An efficient and near linear scaling pair natural orbital based local coupled cluster method},
  year      = {2013},
  issn      = {1089-7690},
  month     = jan,
  number    = {3},
  pages     = {034106},
  volume    = {138},
  doi       = {10.1063/1.4773581},
  publisher = {AIP Publishing},
}

@Article{Riplinger2013a,
  author    = {Riplinger, Christoph and Sandhoefer, Barbara and Hansen, Andreas and Neese, Frank},
  journal   = {J. Chem. Phys.},
  title     = {Natural triple excitations in local coupled cluster calculations with pair natural orbitals},
  year      = {2013},
  issn      = {1089-7690},
  month     = oct,
  number    = {13},
  pages     = {134101},
  volume    = {139},
  doi       = {10.1063/1.4821834},
  publisher = {AIP Publishing},
}

@Article{Sheldon2024,
  author    = {Sheldon, Christopher and Paier, Joachim and Usvyat, Denis and Sauer, Joachim},
  journal   = {J. Chem. Theory Comput.},
  title     = {Hybrid {RPA:DFT} Approach for Adsorption on Transition Metal Surfaces: Methane and Ethane on Platinum (111)},
  year      = {2024},
  issn      = {1549-9626},
  month     = feb,
  number    = {5},
  pages     = {2219--2227},
  volume    = {20},
  doi       = {10.1021/acs.jctc.3c01308},
  publisher = {American Chemical Society (ACS)},
}

@Misc{Stevenson2025,
  author       = {Stevenson, James M. and Agarwal, Garvit and Bhai, Lakshmi and Debnath, Sibali and Goldsmith, Zachary K. and Mondal, Monosij and Yeu, Inwon and Guha, Rishabh D. and Svirinovsky-Arbeli, Asya and Steingart, Daniel and Marbella, Lauren and Urban, Alexander and Jacobson, Leif D. and Halls, Mathew David and Friesner, Richard A.},
  howpublished = {chemrxiv-2025-ndqzk-v2},
  month        = sep,
  title        = {Evidence for significant multi-{Li}+ clustering in common lithium-ion battery electrolytes},
  year         = {2025},
  doi          = {10.26434/chemrxiv-2025-ndqzk-v2},
  publisher    = {American Chemical Society (ACS)},
}

@Misc{Wei2026,
  author       = {Wei, Yujing and L. Weber, John and M. Stevenson, James and K. Goldsmith, Zachary and Xie, Xiaowei and D. Jacobson, Leif and A. Friesner, Richard},
  howpublished = {chemrxiv-2025-x1km2},
  month        = feb,
  title        = {An Accurate Charge-Aware Machine-Learning Interatomic Potential for the Reduction of {Li}-ion Battery Electrolytes in Solution},
  year         = {2026},
  doi          = {10.26434/chemrxiv-2025-x1km2/v2},
  publisher    = {American Chemical Society (ACS)},
}

@Misc{Kundu2025,
  author       = {Kundu, Sohang and Chamaki, Diana and Ye, Hong-Zhou and Agarwal, Garvit and Berkelbach, Timothy C.},
  howpublished = {arXiv:2509.14067},
  title        = {Reaction dynamics of lithium-mediated electrolyte decomposition using machine learning potentials},
  year         = {2025},
  copyright    = {arXiv.org perpetual, non-exclusive license},
  doi          = {10.48550/ARXIV.2509.14067},
  publisher    = {arXiv},
}

\end{document}


\title{
Supplementary Material: Adsorption energies and decomposition barrier heights for ethylene carbonate on the surface of lithium from cluster-based quantum chemistry
}

\author{Ethan A. Vo}
\author{Hung T. Vuong}
\author{Zachary K. Goldsmith}
\affiliation{Department of Chemistry, Columbia University, New York, NY 10027 USA}
\author{Hong-Zhou Ye}
\affiliation{Department of Chemistry and Biochemistry, University of Maryland, College Park, MD, 20742, USA}
\affiliation{Institute for Physical Science and Technology, University of Maryland, College Park, MD, 20742, USA}
\author{Yujing Wei}
\author{Sohang Kundu}
\author{Ardavan Farahvash}
\affiliation{Department of Chemistry, Columbia University, New York, NY 10027 USA}
\author{Garvit Agarwal}
\affiliation{Schr{\"o}dinger, Inc., New York, NY 10036, USA}
\author{Richard A. Friesner}
\email{raf8@columbia.edu}
\affiliation{Department of Chemistry, Columbia University, New York, NY 10027 USA}
\author{Timothy C. Berkelbach}
\email{t.berkelbach@columbia.edu}
\affiliation{Department of Chemistry, Columbia University, New York, NY 10027 USA}
\affiliation{Initiative for Computational Catalysis, Flatiron Institute, New York, NY 10010 USA}

\maketitle

\section{Comparison between hemispherical clusters and parallelepiped clusters}

\begin{figure}[h]
    \centering
    \includegraphics[scale=1.0]{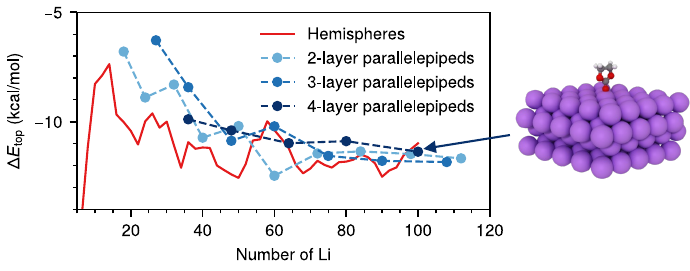}
    \caption{
    Top adsorption energy, calculated with PBE/cc-pVTZ, comparing parallelepiped clusters with the hemispherical clusters used in the main text. 
    }
    \label{fig:prism}
\end{figure}

In Fig.~\ref{fig:prism}, we compare the top adsorption energy calculated using PBE/cc-pVTZ when using parallelepiped clusters versus the hemispherical clusters used in the main text.
The convergence is not noticably smoother with either cluster shape, but the parallelepipeds give access to fewer clusters.

\section{Comparison between canonical and DLPNO-CC}

\begin{figure}[h]
    \centering
    \includegraphics[scale=1.0]{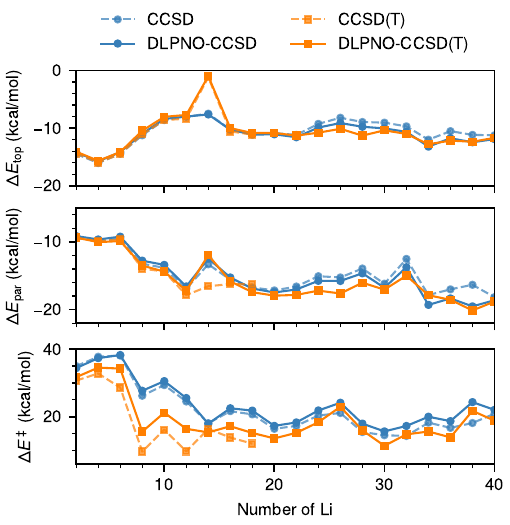}
    \caption{
    Assessment of the DLPNO approximation in the cc-pVTZ basis set.
    }
    \label{fig:dlpno}
\end{figure}

As discussed in the main text, we assessed the domain-based localized pair natural orbital (DLPNO) approximation~\cite{Riplinger2013,Riplinger2013a} in ORCA~\cite{Neese2011,Neese2025} by comparing DLPNO-CCSD and DLPNO-CCSD(T) to their canonical counterparts on clusters containing about 20--40 lithium atoms.
Results are shown in Fig.~\ref{fig:dlpno}.
The approximation is seen to be quite good for CCSD, although the errors increase with increasing cluster size, up to 2--3~kcal/mol for the largest clusters.
The approximation is slightly worse for CCSD(T), especially for the barrier height.

\section{Additional ph-AFQMC calculation details}
DFT trial wave functions are generated using PySCF~\cite{Sun2017,Sun2020} with density fitting. 
All ph-AFQMC calculations utilize a frozen core scheme in which the core 1s orbitals of C and O are frozen, leading to a total of 6 frozen core orbitals for the EC molecule.
This is also the default frozen core scheme employed in ORCA calculations.
We use a modified Cholesky decomposition of the electron repulsion integral tensor with a threshold of $10^{-5}$~a.u.
The localized orbital cutoff for ph-AFQMC calculation is $10^{-4}$ a.u.~\cite{Weber2022}
A total of 2048 walkers are propagated for $250$ Ha$^{-1}$, with a propagation time step of $0.005~\mathrm{Ha}^{-1}$.
We apply population control using the Comb algorithm every $0.1~\mathrm{Ha}^{-1}$.
Walkers are orthonormalized every $0.01~\mathrm{Ha}^{-1}$ for numerical stability, and the local energy is measured every $0.1~\mathrm{Ha}^{-1}$.

As described in the main text, the parallel adsorption energy of four systems was recomputed using an HCI trial wavefunction.
HCI trial wave functions are generated in a modified protocol similar to the ``AFQMC 0'' protocol from Ref.~\citenum{Wei2024}, described as follows.
First, a single-shot HCI calculation is performed with the ``valence mapping" (See Table S4 in Section 3 of the Supporting Information of Ref.~\citenum{Wei2024}) with the parameter $\epsilon_1=10^{-3}$.
A second active space is then chosen by selecting the natural orbitals from the HCI wave function with occupation numbers between $0.02$ and $1.98$, which is then used to initialized an HCISCF calculation with $\epsilon_1=10^{-3}$.
Finally, a single shot HCI calculation with $\epsilon_1=10^{-4}$ is performed with the optimized orbitals to form the final wave function. 
For these four systems, the active space information, as well as the trial wave function and AFQMC energies are given in Table~\ref{tab:si01}.

\begin{table}[H]
\centering
\begin{tabular*}{0.8\columnwidth}{@{\extracolsep{\fill}} lcccccc}

\toprule													
System	&	Number of Li	&	Active space	&	$N_{\mathrm{det}}$ [\%CI]	&	HCI $E$	&	AFQMC $E$	&	AFQMC Stoc. Err.	\\
\midrule													
EC	&	26	&	(2e, 2o)	&	2 [1.00]	&	-340.6007	&	-341.9075	&	0.0006	\\
	&	28	&	(2e, 2o)	&	2 [1.00]	&	-340.6008	&	-341.9087	&	0.0006	\\
	&	38	&	(2e, 2o)	&	2 [1.00]	&	-340.6009	&	-341.9089	&	0.0007	\\
	&	40	&	(2e, 2o)	&	2 [1.00]	&	-340.6009	&	-341.9087	&	0.0005	\\
\midrule													
Li cluster	&	26	&	(20e, 19o)	&	6834 [0.980]	&	-193.6313	&	-194.6574	&	0.0006	\\
	&	28	&	(20e, 19o)	&	8159 [0.985]	&	-208.5163	&	-209.6441	&	0.0009	\\
	&	38	&	(24e, 22o)	&	8434 [0.975]	&	-282.9950	&	-284.5903	&	0.0011	\\
	&	40	&	(22e, 21o)	&	4556 [0.960]	&	-297.8719	&	-299.5694	&	0.0008	\\
\midrule													
EC+Li cluster	&	26	&	(20e, 18o)	&	5033 [0.980]	&	-534.2335	&	-536.5941	&	0.0008	\\
	&	28	&	(20e, 17o)	&	4888 [0.985]	&	-549.1083	&	-551.5748	&	0.0011	\\
	&	38	&	(22e, 20o)	&	5989 [0.975]	&	-623.5960	&	-626.5275	&	0.0013	\\
	&	40	&	(20e, 19o)	&	2764 [0.960]	&	-638.4801	&	-641.5116	&	0.0016	\\
\bottomrule													

\end{tabular*}

\caption{Active space of the HCI trial wave functions, the number of determinants
and corresponding CI percent retained in the AFQMC calculations, HCI and AFQMC total energy, and AFQMC stochastic error.
All calculations are performed in the cc-pVTZ basis and all energy values are reported in Ha.}
\label{tab:si01}
\end{table}

\bibliography{cluster,afqmc_refs}